\documentclass[10pt,letterpaper]{article}
\usepackage{opex3}
\usepackage{color}
\usepackage{amsmath}    
\usepackage{cite}
\usepackage{hyperref}

\begin{document}

\title{Subsampled Phase Retrieval for On-chip Lensless Holographic Video}

\author{Donghun Ryu,$^{\textrm{1,$\dagger$}}$ Zihao Wang$^{\textrm{2,$\dagger$}}$, Kuan He$^{\textrm{2,$\dagger$}}$, Roarke Horstmeyer$^{\textrm{3,4,*}}$ and Oliver Cossairt$^{\textrm{2}}$}
\address{$^{\textrm{1}}$Electrical Engineering, California Institute of Technology, Pasadena, CA 91125, USA \\ $^{\textrm{2}}$Electrical Engineering and Computer Science, Northwestern University, Evanston, IL 60208, USA\\ $^{\textrm{3}}$Charit\'e Medical School, Humboldt University of Berlin, Berlin 10117, Germany\\
$^{\textrm{4}}$Future address: Biomedical Engineering, Duke University, Durham, NC 27708, USA\\
$^{\textrm{$\dagger$}}$These authors contributed equally to this work}
\email{$^{\textrm{*}}$roarke@charite.de} 



\begin{abstract}On-chip holographic video is a convenient way to monitor biological samples simultaneously at high spatial resolution and over a wide field-of-view. However, due to the limited readout rate of digital detector arrays, one often faces a tradeoff between the per-frame pixel count and frame rate of the captured video. In this report, we propose a subsampled phase retrieval (SPR) algorithm to overcome the spatial-temporal trade-off in holographic video. Compared to traditional phase retrieval approaches, our SPR algorithm  uses over an order of magnitude less pixel measurements while maintaining suitable reconstruction quality. We use an on-chip holographic video setup with pixel sub-sampling to experimentally demonstrate a factor of $5.5$ increase in sensor frame rate while monitoring the \emph{in vivo} movement of \emph{Peranema} microorganisms. 
\end{abstract}



\section{Introduction}

On-chip lensless imaging offers the ability to simultaneously obtain a high resolution image over a wide field-of-view (FOV) in a simple optical setup. In an on-chip lensless imaging experiment, one typically places a sample within several millimeters of a digital sensor. By illuminating the sample with a spatially coherent light source, a diffraction pattern is formed at the nearby sensor, which may be captured as a digital in-line hologram. A phase retrieval algorithm typically recovers the sample's amplitude and phase, at high fidelity, from the recorded hologram intensities~\cite{Tseng10}. 

Lensless holographic imaging has been widely used to investigate biological and chemical phenomena at the micro and/or nano scale. Recent examples include high resolution and wide field-of-view imaging of malaria-infected cells~\cite{Bishara11}, dense pathology slides~\cite{Greenbaum14}, and nanometer-scale viruses~\cite{McLeod15}. While these samples are primarily stationary over time, it is also possible to monitor in-vivo dynamic phenomena using lensless holographic video. On-chip examples of monitoring biophysical processes include discovering the spiral trajectories of sperm~\cite{Su12}, the formation of endothelial cells into microvessels~\cite{Weidling12}, and analyzing single-cell motility~\cite{Pushkarsky14}.

The total pixel count of lensless images (i.e., the system space-bandwidth product) is simply set by the effective pixel count of the detector array. As detector array sizes grow into the regime of hundreds of megapixels, a limited detector array readout rate will eventually limit the rate of high-speed lensless image acquisition. A tradeoff space thus emerges between the spatial and temporal resolution of a lensless imaging experiment: either images can be acquired at either high resolution, or at high frame rates, but currently not both. 

The same tradeoff space also currently impacts video capture in conventional cameras. The limited speed of sensor hardware for pixel readout, analog-to-digital conversion, and a constrained on-board memory together form a data bottleneck. To overcome this limitation, many high speed camera sensors now offer a multitude of video frame rates at different image resolutions. A typical example is the recent Casio EX-F1 camera, which trades off image resolution and frame rate in an inversely proportional manner, offering 2.07 megapixels (MP) at 30 frames per second (fps), 0.20 MP at 300 fps, 0.08 MP at 600 fps, and 0.03 MP at 1200 fps~\cite{casio}. Here, the sensor data rate faces an approximate upper bound of 65 MP per second. 

A number of different coding strategies were recently proposed to overcome this data readout limit. For example, offsetting the exposure time of interleaved pixels may simultaneously provide high-speed video and high-resolution imaging~\cite{Bub10}. A similar strategy may be applied to interleaved frames from a camera array~\cite{Agrawal10}. Alternatively, the incident light may be coded into a spatio-temporal pattern, either using a spatial light modulator~\cite{Reddy11,Liu14}, global shutter~\cite{Holloway12} or translating mask~\cite{Llull13}. Subsequently, an inversion algorithm, typically operating within a compressive sensing framework that assumes scene sparsity, can recover a high-resolution and high-speed video\cite{Wang16}. This strategy was most recently applied with a streak camera to create videos of light propagation resolved down to picosecond time scales~\cite{Gao14}.

Similar coding strategies may also help overcome the space-time resolution tradeoff in lensless holographic imaging. Unlike traditional video, however, the operation of a lensless holographic setup is fundamentally connected to its phase-retrieval algorithm. An ideal strategy to improve lensless image readout rates would operate in tandem with phase retrieval. As with the compressive video recovery schemes above, phase retrieval must also assume some prior knowledge about the imaged sample to ensure accurate algorithm convergence. Examples include a known finite sample support~\cite{Fienup82}, sparsity\cite{Song2016}, non-negativity or an intensity histogram. Several recent works examine how sample sparsity permits accurate sample reconstruction from a limited number of holographic measurements~\cite{Brady09,Denis09,He15,Wang16,Hahn11,Szameit12,Gazit09,Rivenson10,Rivenson15,Chan08}. To the best of our knowledge, no work has yet examined whether prior knowledge of sample support alone may also relax required in-line holographic image readout rates, nor has demonstrated tha such a modified phase retrieval process can improve the frame rate of on-chip holographic video.  

Here, we present a simple lensless imaging method and associated sub-sampled phase retrieval (SPR) algorithm that aims to simultaneously offer high resolution over both space and time. Or approach selectively reads off a limited subset of pixels per image frame. This reduces our per-frame data output, which equivalently increases the imaging system's achievable video frame rate, assuming a fixed sensor readout rate. We then recover accurate, high-resolution maps of sample amplitude and phase using just our sparse set of measured intensities, along with a bootstrapped estimate of the sample support or a well-known algorithm to recover the sampple support called shrink-wrap~\cite{Marchesini03}. We demonstrate how this subsampling strategy can reduce the number of measured pixels in each image frame by up to a factor of 30 with minimal impact upon image fidelity (less than a doubling in recovery error) for several realistic objects. We additionally show that the SPR algorithm offers a factor of 5-6 experimental speed-up in video frame in an \emph{in vivo} on-chip experiment.

Here is an outline for the rest of this paper. First, we review the process of phase retrieval for in-line holography. Second, we introduce our proposed sub-sampling strategy. Third, we test our new measurement and reconstruction method in simulation. We show that our subsampled phase retrieval (SPR) technique outperforms the naive approach of image interpolation. Fourth, we demonstrate the successful operation of SPR in two on-chip imaging experiments. We experimentally verify the ability to measure quantitatively accurate sample phase after reducing the number of samples per image by a factor of 25. We then test SPR with an in-vivo imaging experiment, demonstrating a 9X reduction in sampling requirements while imaging motile peranema protists.

\begin{figure}[]
\centering
\includegraphics[width=.7\columnwidth]{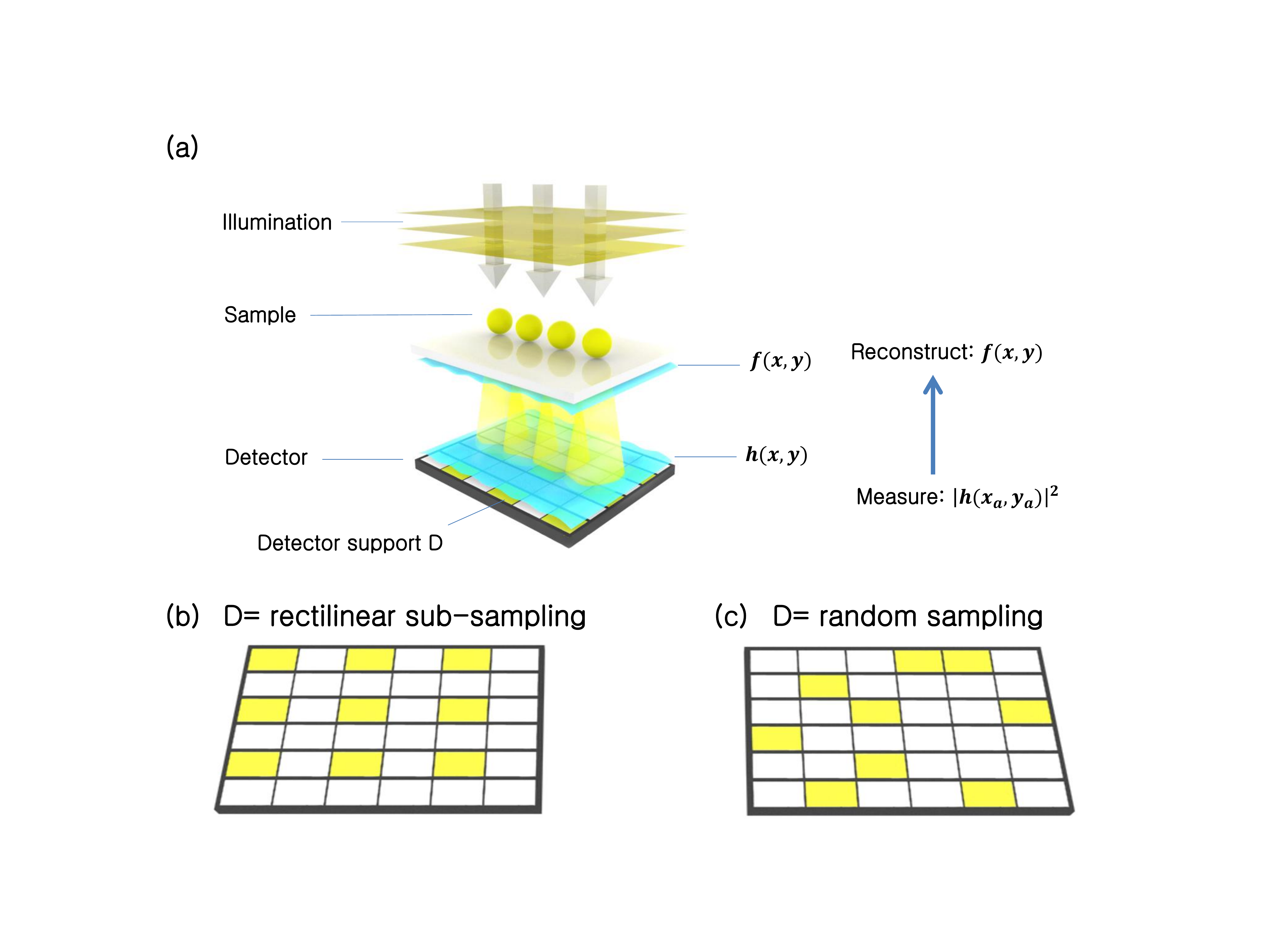}
\caption[2D]{Subsampling phase retrieval (SPR) setup. (a) A detector array images the in-line hologram of a nearby complex sample, $f(x,y)$. The computational goal of ``standard" lensless holography is to determine the complex sample $f$ given the measured hologram amplitudes, $|h|$, from every pixel int the detector array. SPR measures only a subset of values from the detector array (denoted in yellow), selected from (b) a spaced rectilinear grid or (c) randomly. }
\vspace{-0.2cm}
\label{setup}
\end{figure} 
\section{Background and theory}

\subsection{In-line Holography}

A simple schematic of an in-line holography setup is shown in Fig.~\ref{setup}(a). Here, we assume a distant point source illuminates a thin sample with a quasi-monochromatic, spatially coherent plane wave. While not done so here, it is direct to take into account the effects of partially coherent sample illumination~\cite{Mudanyali10}. The optical field immediately after the sample, $f(x,y)=A(x,y)e^{i\phi(x,y)}$, offers a direct indication of its absorptivity within its amplitude $A(x,y)$, and optical thickness within its phase $e^{i\phi(x,y)}$.

The sample field $f(x,y)$ then propagates a distance $d$ to the detector plane, which contains an array of pixels. Directly above this plane, we denote the resulting complex field as $h(x_a,y_a)$, the hologram field, where $(x_a,y_a)$ are the spatial coordinates at this plane. Given sufficient distance between the sample and detector plane, it is possible to perform holography with a reference beam. In this work, we consider reference-free holographic imaging scenarios, which instead rely upon a phase retrieval algorithm to recover the complex field at the sample plane. In this reference-free phase retrieval, as commonly utilized in Coherent Diffraction Imaging, detected intensity is not necessarily regarded as hologram, but as the intensity of diffraction pattern ~\cite{Fienup78}. We would like to note that others have modeled our on-chip imaging set up as in-line holographic system coupled to phase retrieval algorithm ~\cite{Tseng10,Bishara11,Mudanyali10}.

We may describe the diffraction of the sample field $f$ into the hologram $h$ using a propagation operator, $ P_{d}[\cdot]$. Neglecting evanescent field effects and assuming this propagation is lossless, $P_{d}$ is invertible, and its inverse $ P_{d}^{-1}[\cdot]$ represents time-reversed propagation from the detector plane back to the sample plane. The pixel array at the detector plane only detects the intensity of the hologram field:
\begin{equation}
\left|h(x',y')\right|^2=\left|P_{d}\left[f(x,y)\right]\right|^2, 
\end{equation}
where $(x',y')$ are discretized versions of the detector plane coordinates $(x_a,y_a)$, making $\left|h(x',y')\right|^2$ a discrete function. We may assume the propagation operator $P$ also includes the effects of arbitrary pixel discretization. The goal of phase retrieval is to recover an accurate estimate of the complex sample transmission function, $f(x,y)$ from the measured set of intensities, $\left|h(x',y')\right|^{2}$. 

\begin{figure}[]
\centering
\includegraphics[width=.7\columnwidth]{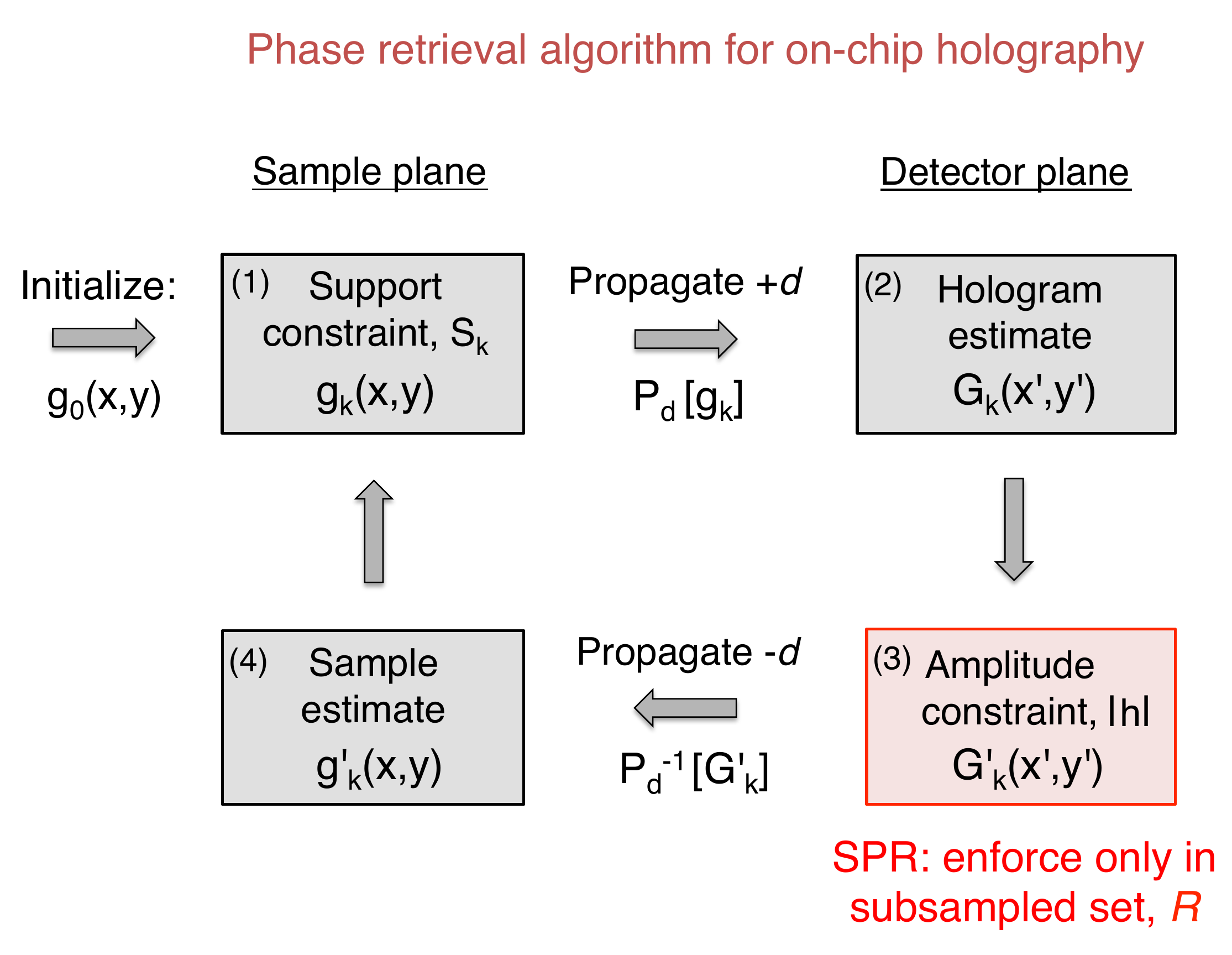}
\caption[2D]{Phase retrieval algorithm for on-chip holography. Each step is detailed in the text. Subsampling (SPR) only modifies the constraint in step 3. This modification results in over an order of magnitude potential speedup for lensless holographic video. }
\vspace{-0.2cm}
\label{algorithm}
\end{figure} 

\subsection{Standard in-line holographic phase retrieval}

Phase retrieval algorithms compute the complex sample field from the measured diffraction pattern intensity through an iterative process~\cite{Fienup82}. Here, we adopt the simple error reduction (ER) algorithm~\cite{Fienup82}. It is also possible to use one of many other closely related strategies~\cite{Marchesini07}, including the hybrid input-output algorithm, or other more advanced solvers~\cite{Kishore15}. Phase retrieval iteratively projects an initial estimate of $f$ onto two constraints in two different domains. In-line holography typically uses for its first constraint the object's support in the sample plane, and for its second constraint the measured hologram intensities in the detector plane.

An outline of the phase retrieval algorithm for this ``standard" case is diagrammed in Fig.~\ref{algorithm}. After initiating an initial complex sample estimate $g_0(x,y)$ at the sample plane, ER first digitally propagates it to the detector plane: $G_k(x',y') = P_{d}[g_k(x,y)]$. Here, $k$ denotes the $k$th iterative loop, for $0 \le k \le n$ iterations. We use capital letters to denote our estimate at the detector plane, and lower case letters to denote it at the sample plane. We perform digital propagation using the angular spectrum method. Next, ER enforces the intensity constraint. It replaces the amplitudes of $G_k(x',y')$ with the experimentally measured amplitudes at the detector, $\left|h(x',y')\right|$:
\begin{equation}
G^{'}_{k}(x',y') = \left|h(x',y')\right| \frac{G_{k}(x',y')}{\left|G_{k}(x',y')\right|}, \forall (x',y') \in D,
\label{amp_constraint1}
\end{equation}
In the standard PR case, $D$ represents the set of all pixels in the detector array and $G^{'}$ is the updated hologram estimate. We may equivalently represent this estimate update as, $G^{'}_{k}(x',y') = \left|h(x',y')\right|e^{i\phi_k(x',y')}$, which makes clear the intensity constraint step leaves the phase of the current hologram estimate, $\phi_k(x',y')$, unchanged. Third, ER propagates this intensity-constrained hologram estimate back to the sample plane: $g'_k(x,y)=P_{d}^{-1}\left[G^{'}_k(x',y')\right]$. Fourth, ER applies a sample support constraint. It leaves unchanged all values within a defined subset of pixels, $S_k$, which typically represents the interior of a collection of cells or an organism of interest. However, it assumes that outside of this interior support area the sample exhibits a uniform absorptivity (i.e., the illumination light primarily passes to the detector unchanged). ER thus sets pixels outside of this support area to a uniform background value $b$: 
\begin{equation}
g_{k+1}(x,y) = \begin{cases} g'_{k}(x,y), & (x,y) \in S_k \\ b, & (x,y) \notin S_k 
\end{cases}
\label{support}
\end{equation}
In this last step the iteration counter value $k$ increments for the next iteration. The above ER loop runs for a fixed number of $n$ iterations, or until some convergence criteria is satisfied. The complex algorithm output, $g_{n}(x,y)$, typically offers an accurate estimate of the amplitude and phase of the original optical field $f(x,y)$ at the sample plane.  

Since we rarely know the exact support of each sample a-priori, we use two recent insights to ensure the constraint in Eq.~\ref{support} encourages successful algorithm convergence. First, we adaptively update the assigned background value, $b$, each iteration. Following~\cite{Mudanyali10}, we set $b=\left< \left|g'_k(x,y)\right| \right> r(x,y)/ \left<r(x,y)\right>$ at iteration $k$, where $\left<\right>$ denotes the mean over all pixels and $r(x,y)$ is a fixed reference measurement formed by back-propagating a set of reference hologram amplitudes, $\left|h_r(x',y')\right|$ (which do not contain diffracted light from any cells or sample structure). The reference amplitudes $\left|h_r\right|$ may be acquired before the experiment, or simply selected from a region of the hologram where no sample structure is present.
 
Second, to improve the accuracy of Eq.~\ref{support}, we also vary the set of pixels defining the sample support each iteration, $S_k$. We update $S_k$ with the ``shrink-wrap" method~\cite{Marchesini03}. At a given iteration, this method first blurs and then thresholds the current sample estimate to form a new support boarder. Blurring helps smooth noise to regularize the support area, and also encourages algorithm stability. Unless otherwise stated, our shrink-wrap implementation uses a Gaussian blur kernel of 5$^{2}$ pixels, a normalized threshold value of 0.15, and updates the support every tenth iteration.  

\begin{figure}[]
\centering
\includegraphics[width=.9\columnwidth]{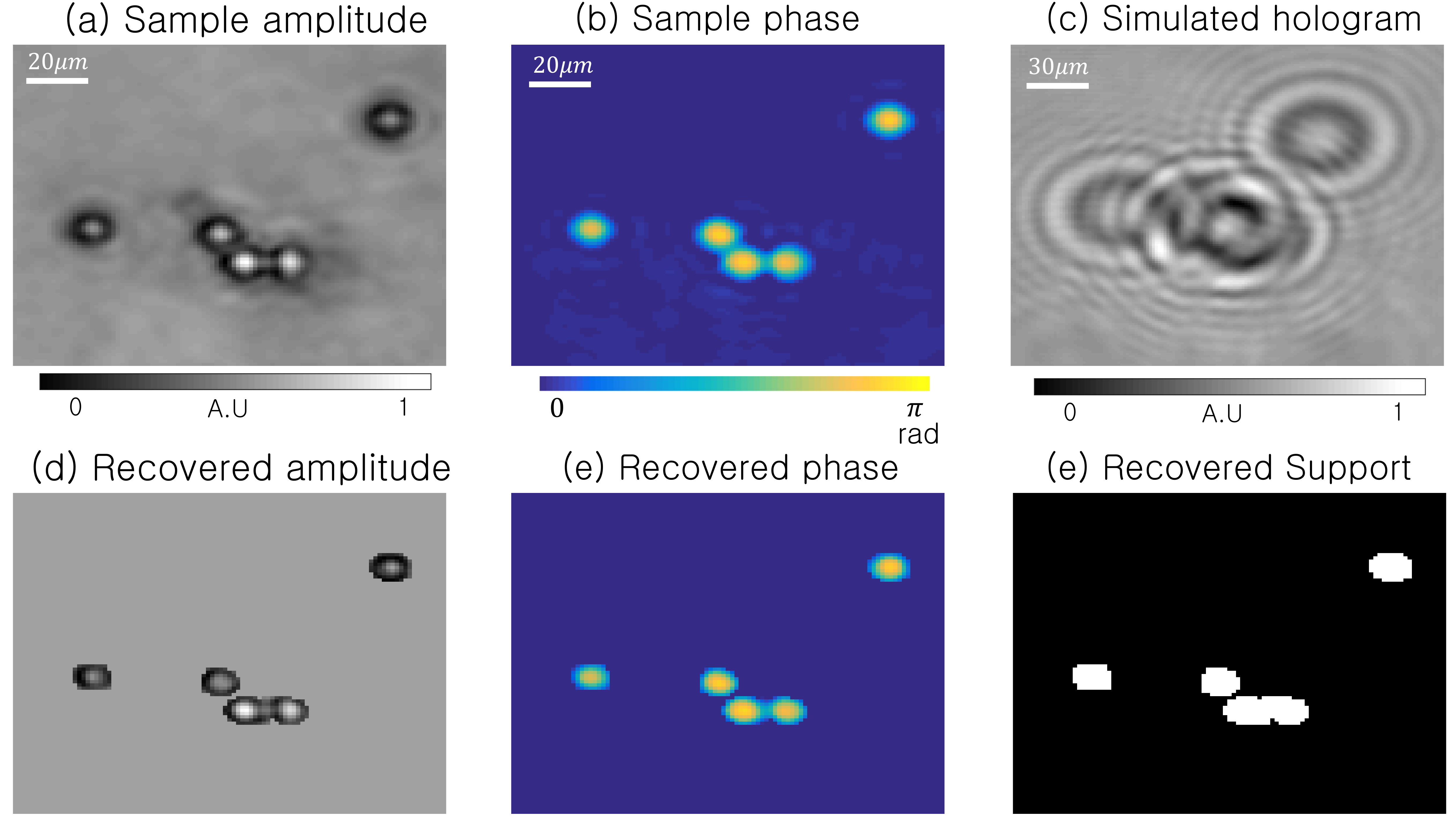}
\caption[2D]{Simulation of standard in-line holography for a set of 5 microspheres. (a)-(b) Simulated sample's original amplitude and phase, leading to (c) detected hologram intensities at sensor plane ($d=1$ mm). (d)-(e) Recovered amplitude and phase using the ER phase retrieval algorithm (updating all pixel amplitudes). (f) The final sample support. }
\vspace{-0.2cm}
\label{standard_sim}
\end{figure}

We show an example simulation of standard ER phase retrieval in Fig.~\ref{standard_sim}. Our simulated sample is $150\times 150$ pixels of measured amplitudes and phases from a set of 5 polystyrene microspheres, shown in Fig.~\ref{standard_sim}(a)-(b), acquired using an alternative phase retrieval approach~\cite{Zheng13,Ou13}. Assuming a lensless imaging setup that approximately matches our experimental parameters (150$^{2}$ pixels, pixel size $= 2.2$ $\mu$m, $d = 1$ mm), we then simulate the formation of a single in-line hologram, which we detect only the intensity of (Fig.~\ref{standard_sim}(c)). From this hologram, we apply the standard ER phase retrieval algorithm, along with shrink wrap support estimation, to recover the complex sample estimate in Fig.~\ref{standard_sim}(d). Fig.~\ref{standard_sim}(f) shows the final sample support. We note that our reconstruction offers quantitatively accurate amplitude and phase {\it within} each microsphere, but sets the complex field to a constant value in all ``background" areas outside of each sphere.  


\subsection{Subsampled phase retrieval (SPR)}
Our SPR algorithm makes one small but critical change to the ER phase retrieval workflow. Instead of measuring the hologram amplitude with all pixels in the digital detector array (the entire set of pixels $D$), SPR uses only a subset of available pixels, $R \subset D$. We replace the estimated hologram amplitude only at these pixel locations, and otherwise leave its estimated complex values unchanged at all other pixel locations. The third step of our SPR algorithm thus takes the form,
\begin{equation}
G^{'}_{k}(x',y') = \begin{cases} \left|h(x',y')\right| \frac{G_{k}(x',y')}{\left|G_{k}(x',y')\right|}, & (x',y') \in R \\ G_{k}(x',y'), & (x',y') \notin R 
\end{cases}
\label{subsampling}
\end{equation}
SPR uses the new constraint in Eq.~\ref{subsampling} for algorithm step 3. All other steps of SPR match the standard ER pipeline. In a digital sensor with a variably addressed pixel readout scheme, SPR only requires $R$ measured intensity values. This reduction in data readout leads to a proportional increase in the detector frame rate. For a given subset size $|R|$, the maximum expected frame rate speedup with SPR is $|D|/|R|$. In this work, we examine multiple data reduction factors ranging from $|D|/|R|=4$ to $|D|/|R|=36$. Furthermore, we investigate two subsampling geometries: rectilinear and random subsampling, as diagrammed in Fig.~\ref{setup}(b)-(c). Rectilinear subsampling periodically skips the readout of a fixed number of pixels along $x$ and $y$ and can currently be achieved by modifying recently available CMOS pixel arrays (details Experiment section). 

We also test SPR using a random subsampling strategy, which helps us directly compare our approach to related method of compressive holography\cite{Brady09,Rivenson10}. Like SPR, the framework of compressive sensing (CS)~\cite{Candes08} can also estimate a complex signal from fewer measurements than originally required by Shannon's sampling theorem. For in-line holography, this offers an alternative means to readout fewer pixels per image, and thus potentially achieve a higher frame rate, while maintaining an accurate reconstruction. CS has two primary requirements. First, the signal must exhibit sparsity - a property requiring that most of the signal energy is contained within just a few coefficients within some transform domain.  Within the context of our lensless imaging setup, compressive sensing might require that our sample exhibit a near-zero amplitude at most pixels in the sample plane, or perhaps exhibit a spatial gradient that is mostly zero (i.e., contains only a few sharp edges). Second, CS requires the sensing matrix and the unknown signal to be incoherent, which is measured by restricted isometric properties~\cite{Candes08b}. Thus, many sparsity-based holography setups rely upon a semi-random sampling strategy, similar to the second subsampling strategy that we consider. We demonstrate the effectiveness of SPR based on a random subsampling scheme both numerically and experimentally.
   
Finally, it is important to note that this subsampling strategy is not equivalent to sampling the hologram with proportionally large pixels and attempting to computationally improve resolution. Larger pixels will not only encounter aliasing issues, but also fail to realize the goal of SPR: to rely more upon the estimated sample support during image reconstruction by leaving a large fraction of hologram values unconstrained. Furthermore, as we will demonstrate, SPR does not equate to image interpolation. The support constraint at the sample plane is vital for filling in both the unknown amplitude and phase of the intermediately empty hologram pixels (i.e., filling in the white pixels in Fig.~\ref{setup}(b)). Due to the sample support constraint, these unknown amplitudes may take on values that are dramatically different from immediately neighboring pixel values, which an image interpolation strategy could not recognize or account for. 
   
Several prior investigations have demonstrated that a holographic measurement, along with a CS reconstruction algorithm, can offer more information than traditional holographic recovery procedures. Examples include recovering the distribution of a sparse sample across three dimensions~\cite{Brady09,Denis09,Hahn11} (i.e., tomographic imaging), and achieving sub-wavelength resolutions within a complex reconstruction from a single complex~\cite{Gazit09} or intensity~\cite{Szameit12} diffraction pattern in the far field. In addition, combining a sparse sample assumption with random subsampling at the detector plane may also lead to accurate hologram recovery~\cite{Rivenson10, Rivenson15}. These same compressive techniques are also useful in the terahertz regime~\cite{Chan08}. While quite similar to the strategy pursued here, no work has yet extended these CS concepts to an on-chip holographic imaging experiment, where it is common to rely upon a finite sample support.

A significant amount of theoretical work now justifies the accuracy of CS recovery techniques. Specifically, one may adopt the proof outlined in \cite{Kishore12a, Kishore12b} to ague that, given a sufficiently sparse sample, phase retrieval from a single image is robust and accurate. Along with insights presented in \cite{Rauhut08}, which are applied within the context of phase retrieval in \cite{Fannjiang11}, this argument may also extend to also account for subsampling at the detector plane. While we provide no formal proof here, this prior work clearly upholds the argument that an appropriate recovery algorithm may accurately determine the amplitude and phase of a sufficiently sparse sample from subsampled in-line hologram measurements.

While SPR does not directly assume the imaged sample is sparse, it does assume a finite spatial support, which we must indirectly acquire. As supported by prior on-chip holography experiments (e.g., \cite{Su12,Weidling12,Pushkarsky14}), we find this support assumption to be realistic for most biological samples of interest. Cells, sperm, embryos and other micro-organisms, for example, all have a well-defined boarder. Given there is a certain amount of overlap in assuming sparsity versus a finite support (that is, both assumptions must set a number of coefficients used to describe the sample to a constant value), we believe that many of the above arguments proving accurate sparse sample recovery might also extend to prove the same for those with finite support. Again, we do not attempt any formal proof of this claim here. Instead, we now demonstrate that SPR is a very effective strategy in practical experiments.

\begin{figure}[]
\centering
\includegraphics[width=.99\columnwidth]{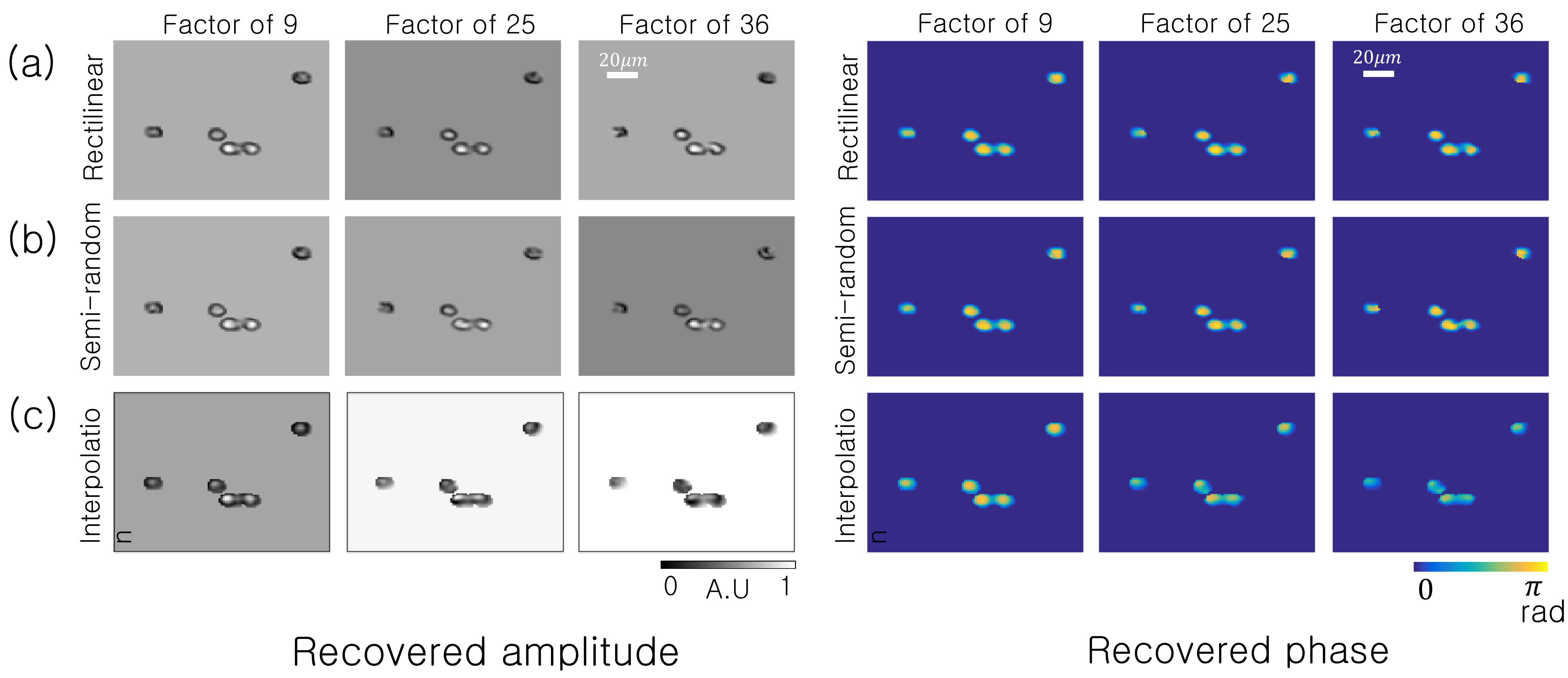}
\caption[2D]{SPR algorithm simulation using the same sample as Fig.~\ref{standard_sim} under two different sampling scenarios: (a) Rectilinear subsampling and (b) random subsampling. For comparison, we also attempt reconstruction after interpolating between our subsampled pixels to form a full-resolution hologram estimate. The results of this interpolation strategy are in (c), where we use the standard ER phase retrieval algorithm for reconstruction. }
\vspace{-0.2cm}
\label{SPR_sim_microspheres}
\end{figure} 

\section{Simulations}

To quantify the effectiveness of SPR in simulation, we assume a digital detector containing 150$^{2}$ pixels (each 2.2 $\mu$m wide) with each sample placed 1 mm above the detector plane (as in Fig.~\ref{standard_sim}). We assume a spatially coherent, quasi-monochromatic source at $\lambda=$632 nm illuminates the sample with a plane wave. Our first simulated sample is the same set of 15 $\mu$m polystyrene microspheres from Fig.~\ref{standard_sim}. Now, instead of constraining our sample estimate with amplitudes measured at all detector pixels, we follow Eq.~\ref{subsampling} and only select a subset $R$ of the detector pixels for three different subsampling ratios: $|D|/|R|=$9, 25 and 36.

The resulting amplitudes and phases of each SPR reconstruction are in Fig.~\ref{SPR_sim_microspheres}. It is clear that one may still faithfully reconstruct the amplitude and phase of this particular sample quite accurately with over an order of magnitude less data per image (i.e., after only reading out values from 1 out of every 36 pixels, either across a grid or randomly). We also compare SPR against the naive approach to subsampling PR: instead of selectively applying the intensity constraint to a subset of pixels, we use image interpolation to infer a ``full-resolution" hologram estimate and apply standard phase retrieval  (i.e., constraining all pixels with Eq.~\ref{amp_constraint1} each iteration). We display the results of this third ``interpolation" strategy in Fig.~\ref{SPR_sim_microspheres}(c), where we apply cubic interpolation to the rectilinearly subsampled intensity data (i.e., to fill in the white pixels in Fig.~\ref{standard_sim}(b) before running the ER algorithm). This exercise highlights that image interpolation appears as a viable strategy, but certain artifacts begin to appear within the reconstructed phase, especially at higher rates of subsampling. 

\begin{figure}[]
\centering
\includegraphics[width=.9\columnwidth]{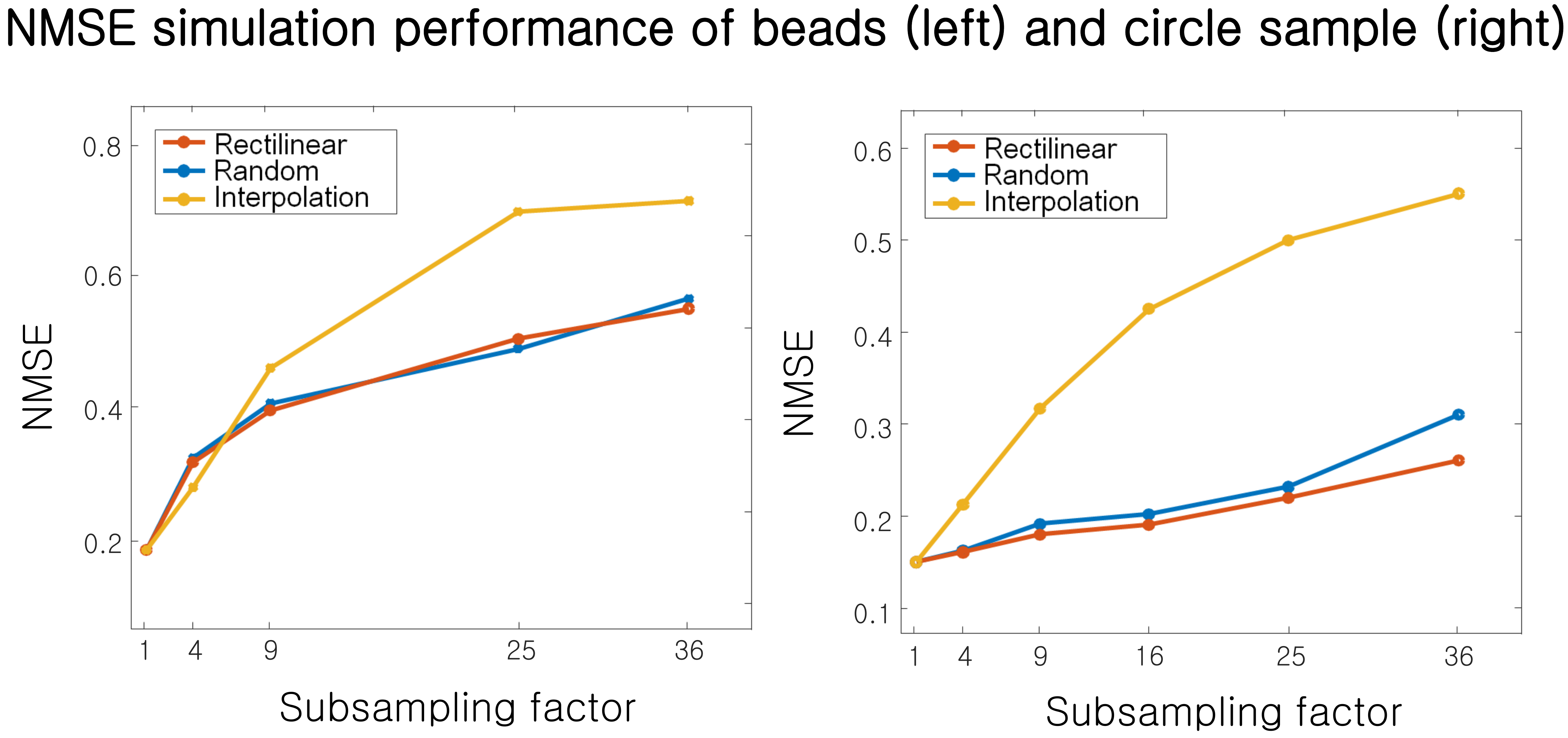}
\caption[2D]{NMSE versus subsampling factor at the detector plane for two simulations: (left) the microscope sample in Fig.~\ref{SPR_sim_microspheres}, and (right) the circular sample in Fig.~\ref{standard_circle}. }
\vspace{-0.2cm}
\label{MSE_plots}
\end{figure} 

To quantitatively compare SPR against the interpolation strategy, we compute the normalized mean-squared error (NMSE) between each reconstruction in Fig.~\ref{SPR_sim_microspheres} and our ground-truth simulation sample, $t(x,y)$. The NMSE metric takes the form:
\begin{equation}
E(k) = \frac{\sum_{(x,y)\in S_n} \left| t(x,y)-\gamma g_n(x,y)\right|^2}{\sum_{(x,y)\in S_n} \left|t(x,y)\right|^2}
\label{NMSE}
\end{equation} 
Here, $g_n(x,y)$ is the recovered sample's amplitude and phase (after $n=500$ iterations) for each of the three strategies outlined in Fig.~\ref{SPR_sim_microspheres} and $S_n$ is the final sample support (no background pixels contribute to our final error metric). The constant parameter $\gamma$ is defined as,
\begin{equation*}
\gamma= \frac{\sum_{(x,y)\in S_n} t(x,y)g_n^{*}(x,y)}{\sum_{(x,y)\in S_n} \left|g_n(x,y)\right|^2},
\end{equation*}
which compensates for the ability of phase retrieval to approach a complex solution only up to an unknown constant phase offset~\cite{Horstmeyer15}.

The microsphere simulation NMSE is in Fig.~\ref{MSE_plots}(a) for subsampling factors ranging from $|D|/|R|=$1 to 36. Both rectilinear and semi-random subsampling offer similar performance. For low amounts of subsampling, the interpolation strategy matches the performance of SPR. However, for high amounts of sub-sampling ($|D|/|R|>9$), SPR has a lower NMSE. After this critical rate, the interpolated values no longer faithfully reproduce the hologram amplitude, and simply leaving the unknown amplitudes unchanged every iteration via SPR becomes a better recovery strategy.  

\begin{figure}[]
\centering
\includegraphics[width=.9\columnwidth]{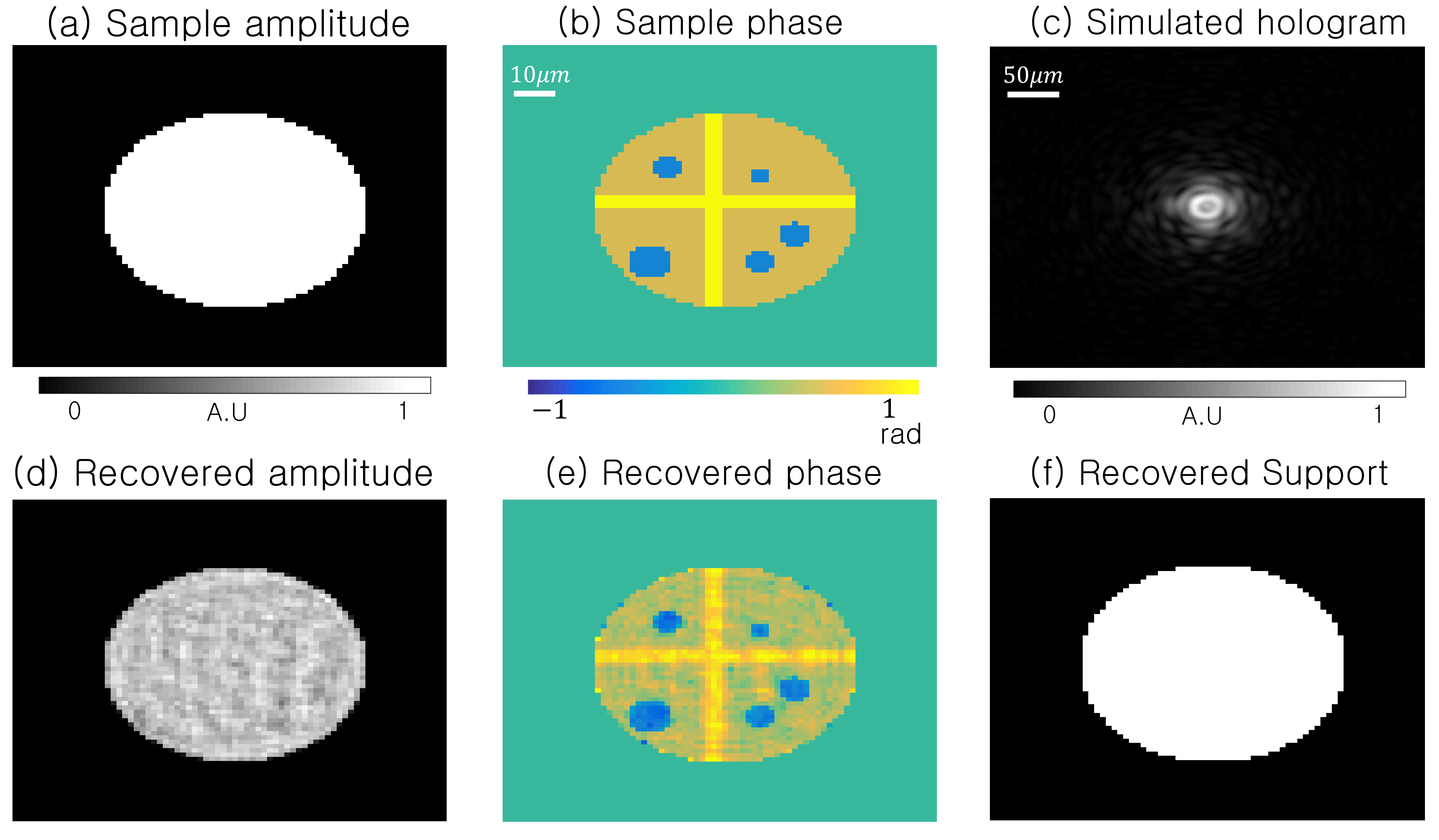}
\caption[2D]{Simulation of standard in-line holography using a circular sample in (a)-(b), with a sharp support boundary. The ER algorithm uses all of the hologram intensities in (c) to recover the sample amplitude (d), phase (e) and support (f). }
\vspace{-0.2cm}
\label{standard_circle}
\end{figure} 

\begin{figure}[]
\centering
\includegraphics[width=.9\columnwidth]{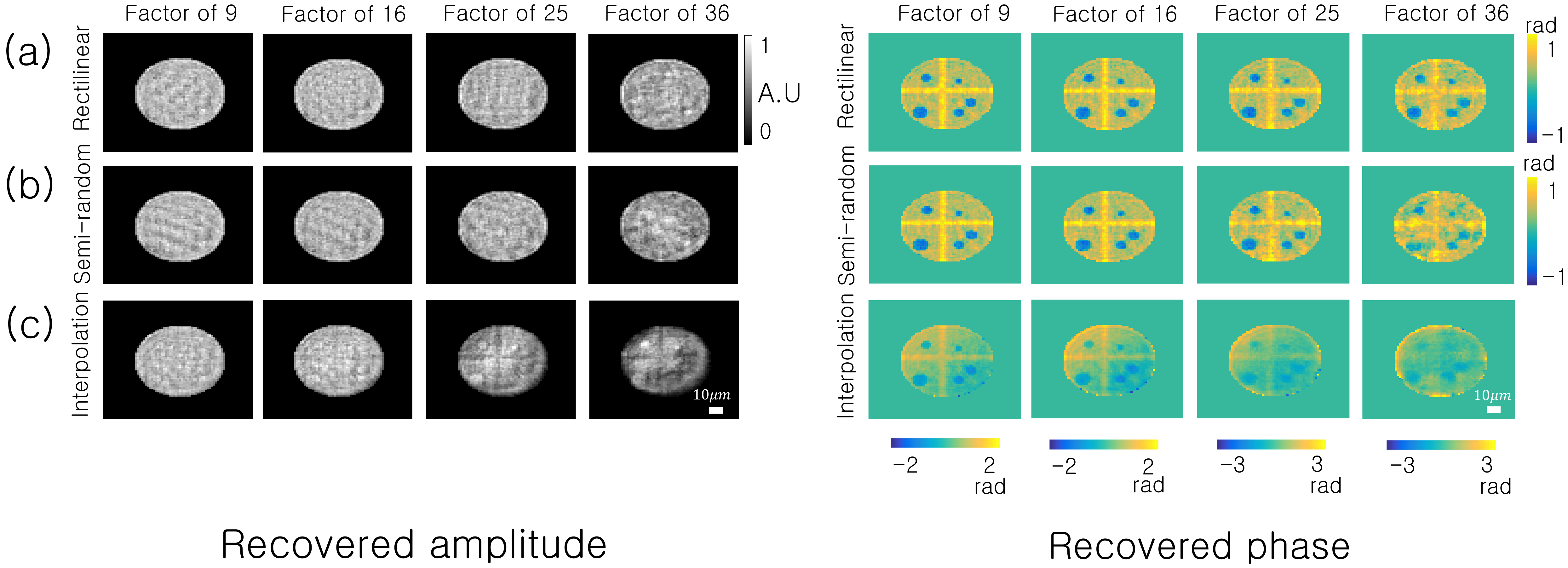}
\caption[2D]{SPR algorithm simulation using the circular sample from Fig.~\ref{standard_circle} under two different scenarios: (a) rectilinear subsampling and (b) random subsampling. For comparison, we also test pixel interpolation after rectilinear subsampling to recover a full-resolution hologram estimate for standard ER phase retrieval. The results of this interpolation strategy are in (c).}
\vspace{-0.2cm}
\label{spr_sim_circle}
\end{figure}

To test the performance of SPR for samples with a sharply delineated support boundary, we attempt a second simulation using the ``circular" sample in Fig.~\ref{spr_sim_circle}. The sample and simulated detector now contain 256$^{2}$ pixels. Again implementing SPR with rectilinear subsampling and semi-random subsampling leads to the reconstructions in Fig.~\ref{spr_sim_circle}(a) and (b), respectively, with the interpolation PR results in Fig.~\ref{spr_sim_circle}(c). This example clearly highlights that SPR easily outperforms interpolation when the sample exhibits a well-defined support boundary. Both the recovered amplitudes and phases are better approximated for all data reduction factors ranging from 1 to 36. As shown in the NMSE plot in Fig.~\ref{MSE_plots}(b), less than 3\% of the original hologram image is required to recover the primary phase features of this particular sample, with only a 2x increase in NMSE. We conclude that, especially for samples exhibiting a well-defined support boundary, SPR can reduce per-image pixel readout by over an order of magnitude with minimal impact reconstruction fidelity.
  
\section{Experiments}

We now present two experimental verifications of SPR. The first experiment use the on-chip imaging setup for hologram capture with an Aptina CMOS sensor (Aptina MT9M002, monochromatic) containing 2560$\times$1920 pixels, each 2.2 $\mu$m wide. The second experiment use the same setup with different sensor (IDS uEyeLE UI-148xLE, monochromatic). This particular sensor now offers a rectilinear sub-sampling for an increase in frame rate. We position each sample a distance $d$ above the active pixel layer of the CMOS detector ($d$ is different in each experiment) and use a red LED (Thorlabs M625L3, 625 nm center wavelength, 16 nm spectral bandwidth) placed 1500 mm above the sample for illumination. To increase the spatial coherence of the LED, we also place a 100 $\mu$m pinhole directly in front of the LED active area. By sweeping many different depths, we manually choose the distance between sample and detector. As for the support constraint, the modified shrink-wrap could successfully yield the support from the measured holograms in all subsampling factors.
%
%

\begin{figure}[]
\centering
\includegraphics[width=.99\columnwidth]{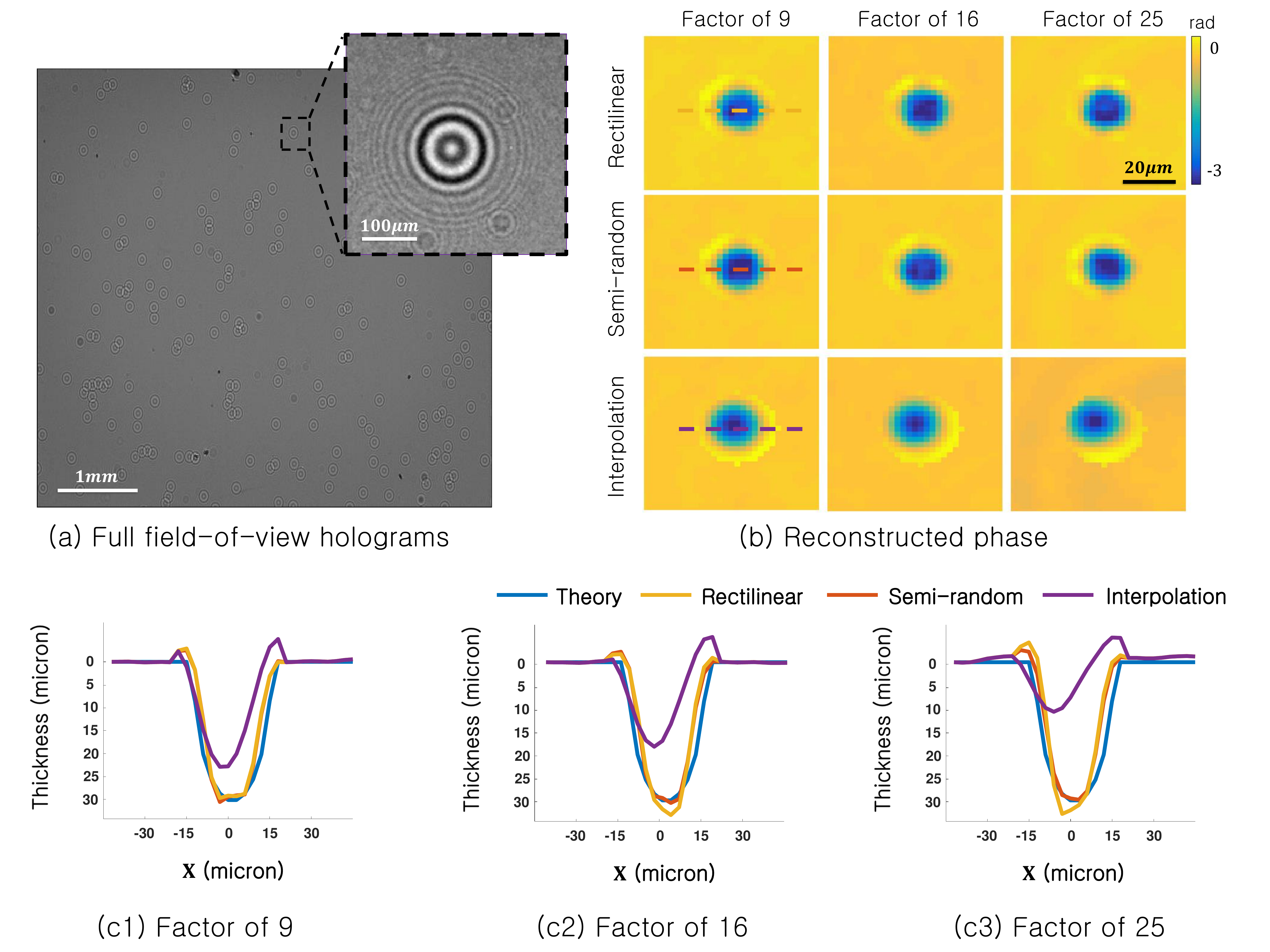}
\caption[2D]{Experimental results, on-chip imaging of polystyrene microspheres. (a) Raw detected hologram with one region of interest highlighted. (b) The recovered sample phase from the region of interest using the SPR algorithm (top and middle) and standard ER phase retrieval with interpolation (bottom). (c) Line traces through the center of the recovered microsphere phase (dashed lines) reveal quantitative agreement with the expected phase shift, even after reducing the number of pixels in factor of 9, 16 and 25. }
\vspace{-0.2cm}
\label{microsphere_exp}
\end{figure} 

First, we verify the ability of SPR to measure quantitative phase by imaging a fixed sample of polystyrene microspheres (30 $\mu$m in diameter, $n_m=1.5875$ refractive index, immersed in oil with $n_o=1.595$ refractive index). We first capture a full-resolution hologram of a large distribution of microspheres. One microsphere of interest, from a 300$\times$300 pixel region, is shown in Fig.~\ref{microsphere_exp}(a). We digitally subsample this measured hologram in two different geometries (rectilinear and semi-random, as outlined in Section 4) at the following subsampling rates: $|D|/|R|=9$,16, and 25. Then, we input these subsampled images into our SPR algorithm and run $n=200$ iterations to recover the microsphere reconstructions shown in Fig.~\ref{microsphere_exp}(b). 

Although up to 96\% of the original hologram image remains unused, SPR still accurately recovers the phase shift induced by each microsphere, as shown for example in Fig.~\ref{microsphere_exp}(b). Here, we also attempt sample reconstruction after first performing cubic image interpolation on the sub-sampled holograms, which results in an unpredictable shift in the phase centroid. Finally, we quantify the accuracy of SPR by comparing its reconstructed phase to the known phase shift induced by an ideal sphere in Fig.~\ref{microsphere_exp}(c1-c3). Here, we select the experimental phase shift values $\Delta\phi$ from along one row of pixels through the center of each sphere (dashed line). The known microsphere phase shift is determined by the optical path length difference of a wave passing through a 30 $\mu$m circle with an index shift of $n_m-n_o=.0075$. From this experiment, we conclude that SPR maintains an accurate measure of quantitative phase. 

\begin{figure}[]
\centering
\includegraphics[width=.99\columnwidth]{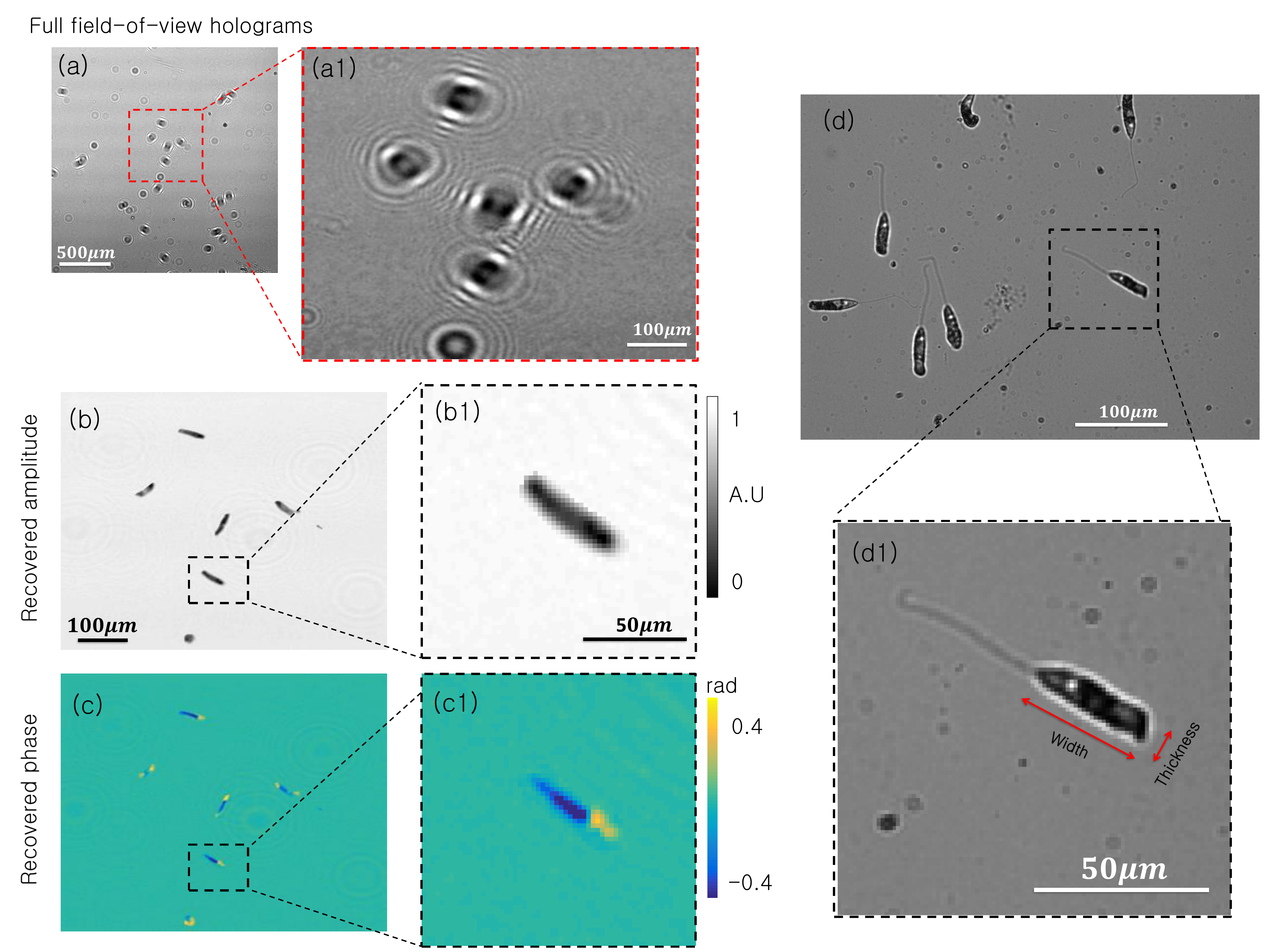}
\caption[2D]{Experimental results for on-chip imaging of live peranema microorganisms. (a) Full field-of-view hologram. (a1) Inset of interest. (b) Reconstructed amplitude with standard ER phase retrieval. (b1) Recovered amplitude of single peranema. (c) Reconstructed phase with standard ER phase retrieval. (c1) Recovered phase of single peranema. (d) 10X microscope image of live peranema. (d1) Inset of single peranema offers a useful ``ground truth" comparison. }
\vspace{-0.2cm}
\label{peranema_exp}
\end{figure}

Given quantitatively accuracy, we next use SPR with an \emph{in vivo} biological specimen. A collection of {\it Peranema}, which are microorganisms that are primarily transparent and fall within the euglinoid family, are placed in medium onto a standard microscope slide. We position the slide 1910 $\mu$m above the sensor and capture a series of holograms, a cropped example of which is shown in Fig.~\ref{peranema_exp}(a). First, we select a 300$^{2}$ pixel region of the hologram and perform standard ER phase retrieval to reconstruct the sample amplitude and phase shown in Fig.~\ref{peranema_exp}(b)-(c). 
 
For ground-truth comparison, we also place the same sample of {\it Peranema} beneath a 10X objective microscope and capture the intensity image in Fig.~\ref{peranema_exp}(d). Although the locations of each microorganism differ from those in reconstruction images due to their unpredictable movement, the structures of their main body qualitatively match. Two other qualitative points are worth noting: first, our reconstructed phase shows a clear boundary between the front and back section of each microorganism, consistent with the presence of their basal body. Second, our reconstructed images cannot resolve the microorganism flagellum (i.e., tail), which is primarily due to our system's limited resolution. We experimentally determined the tail width as approximately 2 $\mu$m, which is close to the 2.2 $\mu$m pixel size of the digital sensor. Future experiments may resolve the flagellum by using a sensor with smaller pixels or with additional processing (see discussion section). 

\begin{figure}[]
\centering
\includegraphics[width=.99\columnwidth]{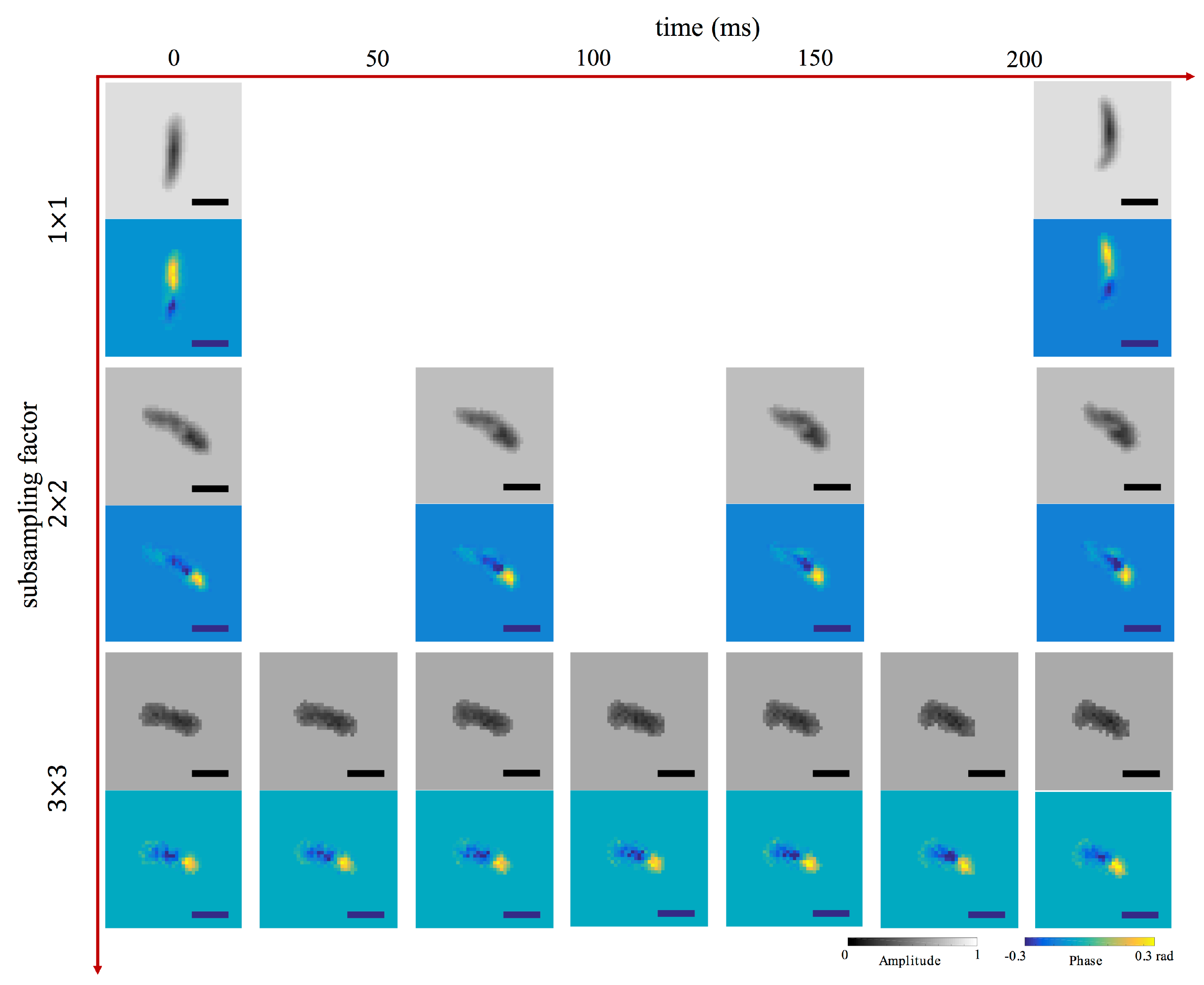}
\caption[2D]{Example: subsampled holographic reconstruction of {\it in vivo Peranema} in motion (subsampling factor vs. time). Horizontal axis depicts time and vertical axis represents subsampling factor. Reconstructions of both amplitude and phase using all pixels on the detector are shown at top, while reconstructions from subsampled pixel array data, using a factor of 4 and 9, are in middle and bottom, respectively. Consecutive frames show {\it Peranema} motion from left to right. Frame rate: first row -- 4.4FPS, second row -- 13.6FPS, third row -- 24.8FPS. Scale bar is 22 $\mu$m. See \textcolor{blue}{Visualization 1, 2 and 3} for the full videos.} 
\vspace{-0.2cm}
\label{peranema_spr}
\end{figure}

Next, we operate the second CMOS sensor which provides rectlinear subsampling mode to capture holographic movies of microorganisms moving over time. We test 3 different subsampling strategies: $|D|/|R|=1$, 4, and 9. For this particular sensor, these subsampled data rates correspond to the ability to increase the sensor frame rate by a factor of 1, 3.1 and 5.5, respectively (from 4.4 FPS for no subsampling to 24.8 FPS for $9\times$ subsampling). For each frame, we apply the SPR algorithm to recover amplitude and phase of each {\it Peranema} across the entire sensor. Example insets of the recovered amplitude and phase of single {\it Peranema} are displayed in Fig. ~\ref{peranema_spr}. Supplemental videos, are provided as \textcolor{blue}{Visualization 1, 2 and 3}, demonstrate how our subsampling strategy offers videos with much smoother motion between consecutive frames, which is not originally captured in full resolution reconstructions. Again, the support from subsampled holograms in each frame could be generated by the modified shrink-wrap algorithm, which plays an important role in the holographic video reconstruction of the {\it Peranema}.

Finally, to quantitatively verify the accuracy of our SPR reconstructions, we compare the average width and thickness of a collection of {\it Peranema} bodies as measured from a single reconstructed frame (with $|D|/|R|=9$) to that measured from a standard microscope image. The mean value (MV) and standard deviation (SD) of the width of the {\it Peranema} from the SPR reconstructed frame is 50.137 $\mu$m and 6.344 $\mu$, respectively, which closely matches that from the microscope image (MV$=47.818$ $\mu$m and SD$=2.511$ $\mu$m, width labeled in Fig.~\ref{peranema_exp}(d1)). Similarly, for the reconstructed thickness we have MV$=11.579$ $\mu$m and SD$=1.607$ $\mu$m for the SPR reconstruction, whereas the microscope image yields MV$=12.937$ $\mu$m and SD$=2.786$ $\mu$m. Both width and thickness match within one standard deviation. 

\section{Discussion and Future Work}

As we demonstrated both in simulation and experiment, SPR can dramatically reduce the number of measurements per frame in on-chip holography while still maintaining suitable reconstruction quality. Using a sensor that achieves a higher frame rate via pixel sub-sampling, we demonstrated a factor of $5.5{\times}$ speedup in holographic video of moving microorganisms.  Our experimental work demonstrates SPR is quite resilient to unknown sensor and shot noise. Placed in the context of alternative compressive holography schemes, SPR is simple, computationally efficient and accurate. 

Several steps may help further improve the accuracy of subsampling. The primary challenge faced with live biological specimen imaging was correctly determining its support, using a somewhat arbitrary starting point. A modified shrink-wrap algorithm, which could incorporate prior knowledge of e.g. the {\it Peranema} body shape, would certainly help improve performance. In addition, the challenge of support identification becomes increasingly difficult when measuring from fewer pixels (i.e., with larger subsampling). Thus, a practical implementation might employ a bootstrapped approach, where the imaging process begins with a larger number of measured pixels per image and then forms a model of the expected sample support to use in later reconstructions with fewer measured pixels. Furthermore, this work offered an initial demonstration of our sampling algorithm on a standard CMOS sensor with a limited ability to modify pixel readout. SPR would ideally benefit from a fully addressable pixel readout scheme. 


We believe SPR offers a useful conceptual starting point for more advanced procedures. First, SPR currently does not consider the redundant nature of the video signal over time. Adopting the insights gained by SPR into a more general approach to optimize phase retrieval over both space and time will likely lead to additional video speedup. Methods such as optical flow may provide a good path forward in this regard. Second, SPR is capable of removing objects that are not in focus, which offers a means to simultaneously achieve optical sectioning. Third, the effectiveness of SPR indicates that it might also be useful for X-ray imaging and coherent diffraction imaging, as well as related techniques for ptychography.
\section*{Funding}
D.R acknowledges Caltech Electrical Engineering department Fellowship. Z.W. and W.K. acknowledge funds in part by NSF CAREER award IIS-1453192, ONR award 1(GG010550)//N00014-14-1-0741, ONR award N00014-15-1-2735, and DARPA award (G001534-7510)//HR0011-16-C-0028. R.H. acknowledges financial support from the Einstein Foundation Berlin.

\section*{Acknowledgements}
We are thankful for the discussions and generous help from Dr. Guoan Zheng, Dr. Xiaoze Ou, Dr. Mooseok Jang and Jaebum Chung. 
We also thank Dr. Changhuei Yang at Caltech and Dr. Do Young Noh at GIST for letting us use their lab equipment. 

\end{document}